\documentclass[aps,amssymb,amsmath,showpacs,twocolumn,prc,preprintnumbers]{revtex4-1}

\usepackage{epsfig}
\usepackage{graphicx}

\newcommand{\zapj}{Astrophys.~J.}
\newcommand{\zapjs}{Astrophys.~J.~Suppl. Ser.}

\begin{document}

\title{Search for new resonant states in $^{10}$C and $^{11}$C and their impact 
on the cosmological lithium problem}

\author{F.~Hammache$^{1}$}\altaffiliation[Corresponding author: ]{hammache@ipno.in2p3.fr}
\author{A.~Coc$^2$}
\author{N.~de S\'er\'eville$^1$}
\author{I.~Stefan$^1$}
\author{P.~Roussel$^1$}
\author{S.~Ancelin$^1$}
\author{M.~Assi\'e$^1$}
\author{L.~Audouin$^1$}
\author{D.~Beaumel$^1$}
\author{S.~Franchoo$^1$}
\author{B.~Fernandez-Dominguez$^3$}
\author{S.~Fox$^4$}
\author{C.~Hamadache$^2$}
\author{J.~Kiener$^2$}
\author{A.~Laird$^4$}
\author{B.~Le Crom$^1$}
\author{A.~Lefebvre-Schuhl$^2$}
\author{L.~Lefebvre$^1$}
\author{I.~Matea$^1$}
\author{A.~Matta$^1$}\altaffiliation[Present address ]{Department of Physics, University of Surrey, Guildford, GU2,5XH, United Kingdom}
\author{G.~Mavilla$^1$}
\author{J.~Mrazek$^5$}
\author{P.~Morfouace$^1$}
\author{F.~de Oliveira Santos$^6$}
\author{A.~Parikh$^7$}
\author{L.~Perrot$^1$}
\author{A.M.~Sanchez-Benitez$^8$}\altaffiliation[Present address ]{GANIL, CEA/DSM-CNRS/IN2P3, Caen, France}
\author{D.~Suzuki$^1$}
\author{V.~Tatischeff$^2$}
\author{P.~Ujic$^6$}\altaffiliation[Present address ]{Vin\u{c}a Institute of Nuclear Sciences, University of Belgrade, Serbia}
\author{M.~Vandebrouck$^1$}

\affiliation{$^1$ Institut de Physique Nucl\'eaire d'Orsay, UMR8608, IN2P3-CNRS, Universit\'e Paris Sud, 91406 Orsay, France} 
\affiliation{$^2$ CSNSM, IN2P3-CNRS,  Universit\'e Paris Sud, 91405 Orsay, France} 
\affiliation{$^3$  Universidade de Santiago de Compostela, E-15786 Santiago, Spain} 
\affiliation{$^4$  Department of Physics, University of York, Heslington, York YO10 5DD, United Kingdom}  
\affiliation{$^5$ Nuclear Physics Institute ASCR, 250 68 Rez, Czech Republic} 
\affiliation{$^6$GANIL, CEA/DSM-CNRS/IN2P3, Caen, France}
\affiliation{$^7$ Departament de F\'isica i Enginyeria Nuclear, Universitat Polit\`ecnica de Catalunya, E-08036 Barcelona, Spain}
\affiliation{$^8$ Departamento de F\'isica Aplicada, Universidad de Huelva, E-21071 Huelva, Spain}

\date{07/11/2013 }

\pacs{ 26.35.+c, 25.55.Kr, 24.30.-v, 29.30.-h, 29.30.-h}

\begin{abstract}

The observed primordial $^7$Li abundance in metal-poor halo stars is found to be lower than its Big-Bang
nucleosynthesis (BBN) calculated value by a factor of approximately three.  Some recent works suggested the possibility 
that this discrepancy originates from missing resonant 
reactions which would destroy the $^7$Be, parent of $^7$Li. The most promising candidate resonances which were found 
include a possibly missed 1$^-$ or 2$^-$ narrow state around 15 MeV in the compound nucleus $^{10}$C formed by 
$^7$Be+$^3$He and a state close to 7.8 MeV in the compound nucleus $^{11}$C formed by $^7$Be+$^4$He. 
In this work, we studied the high excitation energy region of $^{10}$C and the low excitation energy 
region in $^{11}$C via the reactions $^{10}$B($^3$He,t)$^{10}$C and $^{11}$B($^3$He,t)$^{11}$C, 
respectively, at the incident energy of 35 MeV. Our results for $^{10}$C do not support 
$^7$Be+$^3$He as a possible solution for the $^7$Li problem. Concerning $^{11}$C results, the data show no 
new resonances in the excitation energy region of interest and this excludes $^7$Be+$^4$He reaction channel as an explanation for the 
$^7$Li deficit.   

\end{abstract}

\maketitle

The primordial nucleosynthesis of light elements $^2$H, $^{3,4}$He and $^7$Li, 
together with the expansion of the Universe and the cosmic microwave background (CMB) are 
the three observational pillars of the standard Big-Bang model where the last free parameter 
was the baryonic density of the Universe, $\Omega_b$. 
A precise value for this free parameter  has been deduced from the analysis of the anisotropies of the CMB as observed
by the Wilkinson Microwave Anisotropy Probe (WMAP) satellite ($\Omega_b$h$^2$=0.02249$\pm$0.00056) \cite{WMAP}
and more recently by the Planck mission ($\Omega_b$h$^2$=0.02207$\pm$0.00033) \cite{Planck}. 

A comparison between the 
calculations of the primordial abundances of the light nuclei and the observations reveals
a good agreement for helium and an excellent agreement for deuterium.
In contrast, the theoretical predictions show a 
discrepancy by a factor of $\approx3$ for $^7$Li abundance \cite{Cyb08,Coc10}. 
Indeed, at the baryonic density of the Universe, $\Omega_b$h$^2$, derived from the CMB anisotropies, 
the predicted BBN abundance of  $^7$Li is: ($^7$Li/H)$_{BBN}$=(5.12$^{+0.71}_{-0.62}$)$\times$10$^{-10}$ \cite{Cyb08}
when using WMAP data or (4.89$^{+0.41}_{-0.39}$)$\times$10$^{-10}$ \cite{Coc13} with the Planck data. 
On the other hand, the observed $^7$Li abundance, derived from the observations of 
low-metallicity halo dwarf stars, was found to be 
($^7$Li/H$)_{halo*}$ =  $(1.58 \pm 0.31) \times 10^{-10}$ \cite{Sbo10}.
This significant discrepancy between the observations and the BBN predictions is known as the ''lithium problem'' \cite{Lithium2012a}. 

Several ideas were addressed to try to explain this $^7$Li problem \cite{Fie11}. Some conceived the idea that 
the $^7$Li deficit points toward physics beyond the standard model such as decay of super-symmetric particles
 \cite{Fie11}. Others have suggested that the problem could be due to $^7$Li stellar 
destruction in the atmosphere of the halo stars \cite{Ric05}. However, a uniform destruction of $^7$Li over the so-called 
Spite-plateau region seems difficult \cite{Lithium2012b}. Finally, several authors investigated 
the nuclear aspect of the problem concerning the $^7$Li abundance \cite{Coc04, Ang05, Cyb04}. 
The main process for the production of the BBN $^7$Li at $\Omega_b$h$^2$ is the decay of $^7$Be which is 
produced in the reaction $^3$He($^4$He,$\gamma$)$^7$Be. 
Direct measurements of this reaction cross--section were performed by several groups resulting in a significant
reduction of the thermonuclear reaction rate uncertainty \cite{DIL09}, but no solution to the $^7$Li problem. 
More generally, the experimental nuclear data concerning the 12 main BBN reactions  are sufficient to exclude
a solution in this region, so that one has to extend the network to up to now neglected reactions.  
For instance, it was found \cite{Coc04,Cyb12}  that if the $^7$Be(d,p)2$\alpha$ reaction rate was significantly higher,
the $^7$Li  abundance would be brought down to the observed level. However, subsequent  
experiments \cite{Ang05,OMa11,Sch11} ruled out this possibility \cite{Kir11}.

Very recent works have extended this search \cite{CHA10,BRO12,CIV13} and suggested the possibility of partially or totally solving the $^7$Li 
problem if additional destruction of A=7 isotopes occurs via missed resonant nuclear reactions involving $^7$Be. The most 
promising candidates are possibly missed resonant states in $^7$Be+$^3$He$\rightarrow$$^{10}$C \cite{CHA10}, 
and in $^7$Be+$^4$He$\rightarrow$$^{11}$C \cite{BRO12,CIV13} compound nuclei.  

According to Ref. \cite{CHA10}, the presence of any narrow $^{10}$C  states with J$^{\pi}$ =(1$^-$ or 2$^-$), 
allowing s-wave capture, close to the $^7$Be+$^3$He reaction threshold (Q=15.003 MeV), could bring a solution for the $^7$Li problem 
(see Figure 8 in Ref. \cite{CHA10}). Unfortunately,  the excitation energy range of $^{10}$C between 10.0 and 16.5 MeV is very poorly known 
\cite{{SCHN75},{WAN93},{BEN67}}; the investigated excitation energy region in ref \cite{SCHN75}, where $^{10}$C
was populated via $^{10}$B($^3$He,t) reaction, was limited to the region between the ground state and 8 MeV while in 
ref. \cite{WAN93}, where it was studied via (p,n) reaction up to 20 MeV excitation energy, the energy resolution 
of about 1 MeV and the important background doesn't allow to draw any conclusions on the absence or presence 
of a narrow 1$^-$ or 2$^-$ state close to 15 MeV exitation energy. 
$^{10}$C was also studied via $^{12}$C(p,t) reaction but this favors the population of 0$^+$ and 2$^+$ states \cite{BEN67}. 

Concerning the $^7$Be+$^4$He reaction channel, Civitarese et al. \cite{CIV13} claimed also that 
any isolated state between 7.79 and 7.90 MeV excitation energy of $^{11}$C having a total width between 30 and 160 keV 
and corresponding to a resonant energy  between 250 and 350 keV  respectively would solve the $^7$Li problem
(see Figures 5, 6 and 7 in Ref. \cite{CIV13}). The excited states of $^{11}$C up to 9 MeV were studied in the past via various 
indirect reactions \cite{NNDC} and no state between 7.79 and 7.90 MeV is reported. However, no dedicated measurement in 
this narrow energy region was carried out in the previous works, so it can not be excluded that a weakly populated state in this energy region has been missed.  

In the present work, we report on an experimental study of  the excited states of $^{10}$C and $^{11}$C  where 
the high excitation energy region of $^{10}$C and the low one of $^{11}$C were investigated 
using $^{10}$B($^3$He,t)$^{10}$C and $^{11}$B($^3$He,t)$^{11}$C charge-exchange reactions respectively. 
We have chosen to use the ($^3$He,t) reaction because: i) it is less selective than (p,t) reaction where odd 
parity and odd spins are strongly inhibited by the selection rules and
ii) it has much better resolution and detection efficiency than the (p,n) reaction.

The $^{10}$B($^3$He,t)$^{10}$C and $^{11}$B($^3$He,t)$^{11}$C reactions were carried out at an energy of 35 MeV using 
a $^3$He beam from the Tandem accelerator of the Orsay - ALTO facility.
The targets consisted of a  90(9) $\mu$g/cm$^2$ enriched $^{10}$B with a gold
backing of 200 $\mu$g/cm$^2$ thickness and a self-supporting natural boron (80.1$\%$ $^{11}$B) of
250 $\mu$g/cm$^2$ thickness. 
Elastic scattering measurements were performed at 40$^\circ$ in order to evaluate the contaminants 
present in the two targets.  A non-negligible contamination of $^{12}$C and 
$^{16}$O was observed in $^{10}$B and $^{nat}$B targets (Fig.1). Hence, for background contaminant evaluation, 
we performed ($^3$He,t) measurements also on an 80(4) $\mu$g/cm$^2$ $^{12}$C target and a
75(4) $\mu$g/cm$^2$ Si$_2$O$_4$ target. Target thicknesses were determined by $\alpha$-energy loss measurements. 
The beam current was measured with a Faraday cup and the intensity was about 200 enA.  

\begin{figure}[h]
\begin{center}
\includegraphics[width=8.0cm,height=7.cm]{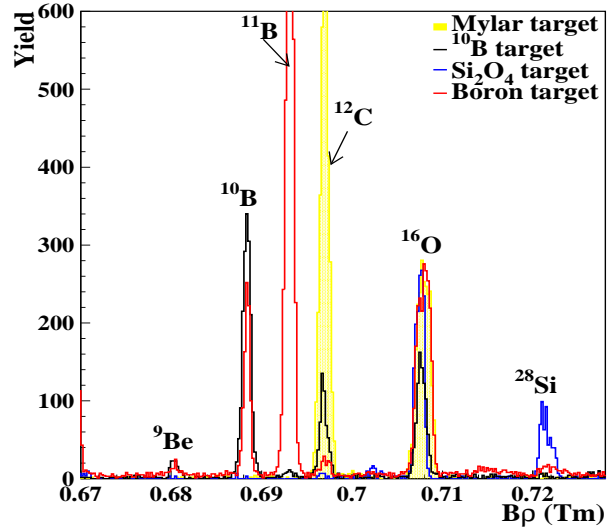}
\caption{(Color online) The black, red, yellow and blue solid lines represent the results from $^3$He
elastic scattering at 35 MeV measured at 40$^\circ$ for $^{10}$B, 
$^{nat}$B, Mylar and Si$_2$O$_4$ targets respectively. The contamination from $^{12}$C and $^{16}$O
can be easily identified in $^{10}$B and $^{nat}$B targets.}
\end{center}
\end{figure}

The reaction products were momentum analyzed by the Split-pole magnetic spectometer \cite{SPEN67} and 
were detected at the focal plane by a position-sensitive gas chamber, a $\Delta$E proportional 
gas-counter and a plastic scintillator. The light fragments emitted by the $^{10}$B($^3$He,t)$^{10}$C 
reaction were detected at 7$^\circ$,  10$^\circ$, 13$^\circ$ and 15$^\circ$ in the 
laboratory system while those coming from $^{11}$B($^3$He,t)$^{11}$C were detected 
at 7$^\circ$ and 10$^\circ$. Well known excited states of $^{11}$C from 2 MeV to 8.655 MeV excitation energy 
were used to calibrate the focal plane position detector. 

The energy resolution obtained in this experiment was about 70 keV (FWHM) in the laboratory frame for the low excitation energy 
region (E$_x$$\leq$8 MeV) and around 37 keV(FWHM) for the high excitation energy region (E$_x$$\geq$12 MeV)
 of $^{10}$C and $^{11}$C. 

Typical spectra obtained for the $^{10}$B($^3$He,t)$^{10}$C reaction at 10$^\circ$ (laboratory) is shown in Fig.2 
for the exitation energy region from 0 to 7 MeV in $^{10}$C 
and in Fig.3  for the excitation region of astrophysical interest from 14 to 16.6 MeV in $^{10}$C. 
Peaks not explicitly labeled in the figures are assigned to unidentified impurities and peaks of $^{10}$C are 
only labelled by their excitation energy.

\begin{figure}[h]
\begin{center}
\includegraphics[width=8.0cm,height=7.0 cm]{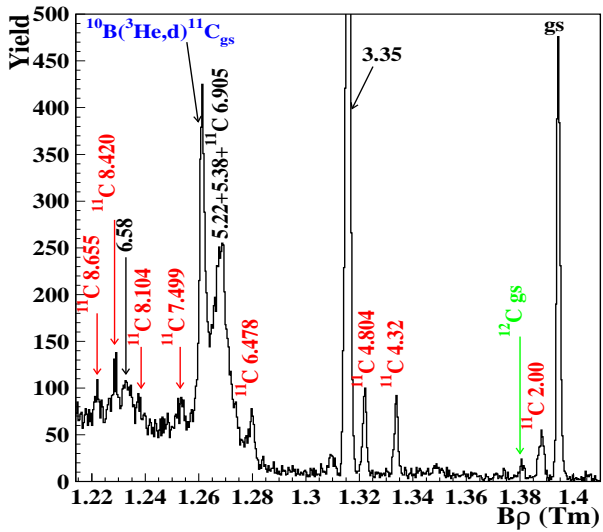}
\caption{(Color online) Triton B$\rho$ spectrum measured at $\theta$=10$^\circ$ (lab) with the 
35 MeV $^3$He beam on the $^{10}$B target in the excitation energy region from gs to 7.2 MeV.
The excitation energy (MeV) of $^{10}$C levels are indicated as well as those of  $^{11}$C coming from a small 
 $^{11}$B contamination of the target or a deuteron contamination of the selected reaction products.}
\end{center}
\end{figure}

\begin{figure}[h]
\begin{center}
\includegraphics[width=8.0cm,height=7.0 cm]{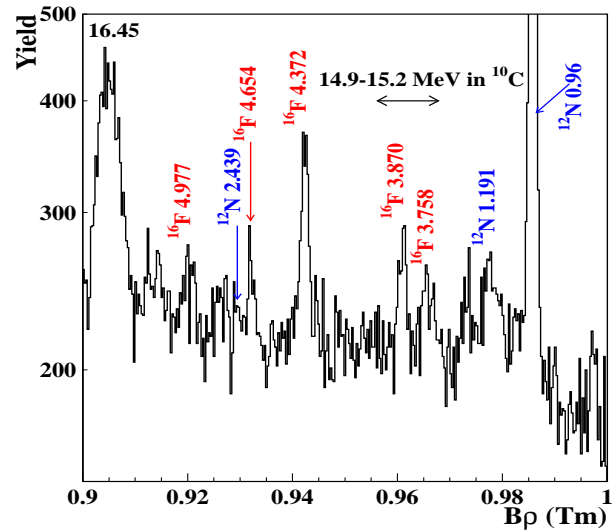}
\caption{(Color online) Triton B$\rho$ spectrum measured at $\theta$=10$^\circ$ (lab) with the 
35 MeV $^3$He beam on $^{10}$B target in the excitation energy region from 14 to 16.5 MeV.
The excitation energy (MeV) of $^{10}$C levels are indicated as well as those of  $^{12}$N and   $^{16}$F
coming from a substantial $^{12}$C and  $^{16}$O contamination of the target. 
The unlabeled peaks correspond to unidentified heavy contamination.}
\end{center}
\end{figure}

The well known levels of $^{10}$C \cite{SCHN75} at  3.35, 5.22, 5.38 and 6.58 MeV excitation energy as well as the ground 
state are observed in this experiment (Fig.2). The 5.22 and 5.38 MeV states are not separated because 
their natural width (225 and 300 keV respectively) is much larger than their separation energy.
A small contamination by the 6.905 MeV state of $^{11}$C due to 
$^{11}$B contamination in the target is also present in the 5.22+5.38 MeV peak.   

The $^{10}$C excitation energy region from 7 to 14 MeV (not displayed in this paper) was also investigated but no other peaks 
except those coming from $^{16}$O($^3$He,t)$^{16}$F and $^{12}$C($^3$He,t)$^{12}$N contaminant reactions 
were identified at the four measured angles. The two states at 9 and 10 MeV excitation energy in $^{10}$C weakly populated in the $^{10}$B(p,n)$^{10}$C 
work \cite{WAN93} are not observed in this experiment. 
 
 In Fig. 3, the background becomes much more important because of the triton 
 continuum. The well populated and separated peaks belong to unbound states of $^{16}$F and bound states of $^{12}$N coming from ($^3$He,t) reactions 
 on $^{16}$O and $^{12}$C contamination, respectively. A dominant peak corresponding to an excitation energy of 16.46 MeV in $^{10}$C is observed at the angles 7$^\circ$ and 10$^\circ$ and follows well the $^{10}$B($^3$He,t)$^{10}$C reaction kinematic. From the peak analysis of this state, we deduced a peak position of 
E$_x$=16.46$\pm$0.01 MeV and a peak width of about  159$\pm$19 keV. 
Note that a state at the energy of about ~16.5 MeV was already observed in the  $^{10}$B(p,n)$^{10}$C experiment \cite{WAN93}. 

Concerning the excitation energy region  between 14.9 and 15.2 MeV in $^{10}$C, no unknown state is observed. 
The only observed peaks in this energy region are the unbound states of $^{16}$F at 3.758 and 3.87 MeV excitation energy due to $^{16}$O contamination in the target. 
The absence of a new $^{10}$C state at the four measured angles may have three non exclusive explanations: first, the state of interest is too large and can not be distuinguished from the triton continuum. Second, the charge exchange ($^3$He,t) reaction cross section is too small, 
or third, there is no isolated state of $^{10}$C in this excitation energy region.

The choice among these explanations is thus dependent on the width of the peak of interest and the ($^3$He,t) cross-section.
In order to visualize the effect of these quantities for our particular experimental conditions a numerical simulation has been
performed of an assumed state at 15.1 MeV on top of the measured ($^3$He,t) spectrum at $\theta_{lab}$=10$^\circ$ once 
the $^{16}$O contamination is substracted from a spectrum obtained with the Si$_2$O$_4$ target. The assumed state is given different widths and is populated with different charge-exchange cross-sections.  
For these simulations, we considered a $^{10}$B thickness of 90 $\mu$g/cm$^2$, a number of incident $^3$He nuclei similar to the 
one measured in the experiment and the measured experimental resolution of 37 keV (FWHM) in the considered excitation energy range. 
The excitation energy region of interest is fitted with a linear function well describing the triton background 
($\chi^2$=0.85 with dof=96, where dof denotes degree of freedom) and the parameters of which are left free. 

The presence of the assumed state with the chosen properties lead to an increase of the $\chi^2$ value, and the null hypothesis "there is a $^{10}$C state in the region of interest'' can be rejected at the usual significance level of 5$\%$ when $\chi^2$ $>$1.25 for 96 degrees of freedom (goodness-of-fit test). An exclusion zone can then be drawn in the plane of the charge-exchange
cross-section versus the width of the assumed state as shown on Fig.4. We conclude that if a state is present in this region, a peak would have 
been identified with a probability larger than 95$\%$ confidence level in the red exclusion zone.

\begin{figure}[h]
\begin{center}
\includegraphics[width=7.6cm,height=6.5 cm]{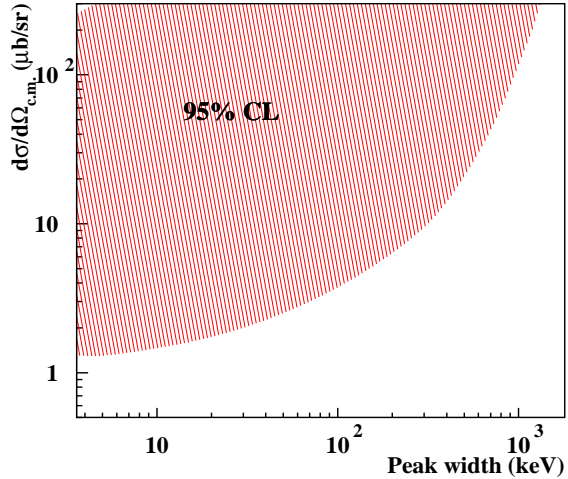}
\caption{[Color online) Exclusion zone in the plane of the charge-exchange cross section vs the width of the state. The red hatched area 
corresponds to a 95$\%$ confidence level (CL) exclusion zone of an isolated $^{10}$C state when using a $^{10}$B target of 90 $\mu$g/cm$^2$ thickness.}
\end{center}
\end{figure}

According to the litterature \cite{{Ball67},{Ball69}},  typical ($^3$He,t) differential charge exchange 
cross sections to 1$^-$ or 2$^-$ states at energies close to our incident energy are generally much 
larger than 25 $\mu$b/sr at 10$^\circ$ in the laboratory system. Therefore, if we assume comparable 
cross sections, we can conclude  from the 95$\%$ CL exclusion zone that any 1$^-$ or 2$^-$ state of $^{10}$C in the excitation energy region around 15 MeV should have, if present, a total width larger than 590 keV in order to not be observed in this experiment. 

To check the consistency of our simulations and conclusions concerning a hypothetical existing $^{10}$C state, 
we used the same procedure for states of $^{12}$N populated by the 
$^{12}$C($^3$He,t)$^{12}$N reaction and using 10 $\mu$g/cm$^2$ of $^{12}$C which is the 
amount of contamination present in our $^{10}$B target. In this case, the total width limit of observation of any 1$^-$ or 2$^-$ existing state of 
$^{12}$N is found to be lower, $\Gamma_T$=300 keV, at the expected cross section of 60 $\mu$b/sr \cite{STE83}. 
This result is in agreement with the observation in our experiment 
of the 1.191 MeV  (J$^\pi$=2$^-$) state of $^{12}$N which has a total width of 
118 keV and the lack of observation of the known 1.8 MeV (J$^\pi$=1$^-$) state of $^{12}$N which has a total width of 750 keV.

Concerning the $^{11}$B($^3$He,t)$^{11}$C measurements,  spectra obtained at 7$^\circ$ (lab) and 10$^\circ$ (lab) 
are shown in Fig. 5 and 6. 

\begin{figure}[h]
\begin{center}
\includegraphics[width=8.cm,height=7.6 cm]{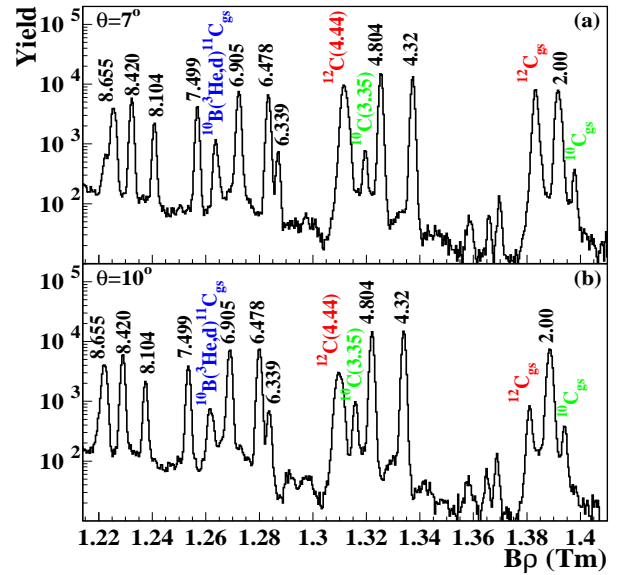}
\caption{(Color online) $^{11}$B($^3$He,t)$^{11}$C B$\rho$ spectra measured at the laboratory angles, $\theta$=7$^\circ$ (a) 
and 10$^\circ$ (b) in the excitation energy region from gs to 9 MeV.  Excitation energies (MeV) of $^{11}$C levels are indicated. 
The unlabeled peaks correspond to unidentified contamination.}
\end{center}
\end{figure}

In Fig.5, we can observe that all the known states of $^{11}$C up to 9 MeV excitation energy are populated with large cross sections in this experiment. 

In Fig.6, we do not observe any triton peak in the excitation energy region between 
7.79 and 7.90 MeV of $^{11}$C corresponding to the energy resonance between 250 and 350 keV. 
Owing to the very large signal to background ratio observed in the spectrum which is ten times better than 
in $^{10}$C case, it is very unlikely that a new state of 
$^{11}$C exists in this energy region. This result is strengthen by the fact that all the known states 
in the mirror nuclei $^{11}$B \cite{NNDC} have their counterpart in $^{11}$C at energies lower than 12 MeV. 

\begin{figure}[h]
\begin{center}
\includegraphics[width=8.cm,height=7.6 cm]{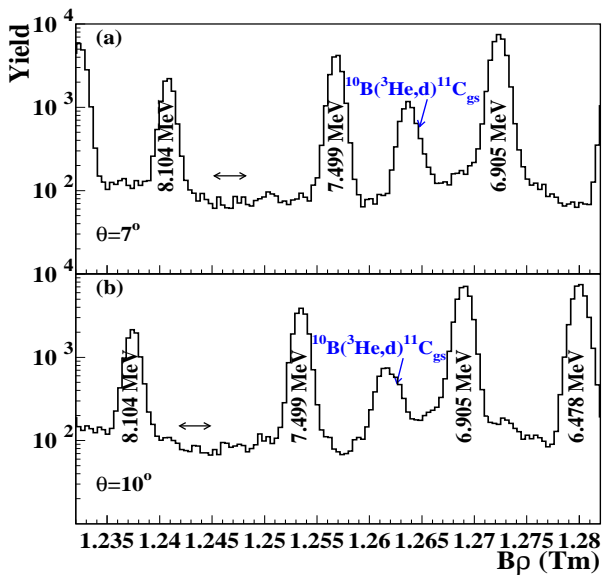}
\caption{(color online) $^{11}$B($^3$He,t)$^{11}$C B$\rho$ spectra measured at $\theta$=7$^\circ$ (a) and 10$^\circ$ (b) 
in the excitation energy region of interest close to 8 MeV.  Excitation energies of $^{11}$C levels are indicated.
The double arrow indicates the astrophysical region of interest.}
\end{center}
\end{figure}

The reactions $^7$Be($^3$He,p)$^9$B and $^7$Be($^3$He,$\alpha$)$^6$Be are the only possible open channels for the reaction
$^7$Be+$^3$He$\rightarrow$$^{10}$C, so calculations of their reaction rates were performed for the lower limit derived from 
our experiment concerning the study of resonant state in the compound nucleus $^{10}$C. 
Hence, we assumed a 1$^-$  state in the compound nucleus $^{10}$C with a total width equal to the experimental lower limit of $\Gamma_{\rm T}$=590 keV in one case and also with $\Gamma_{\rm T}$=200 keV in the case where the ($^3$He,t) differential charge exchange cross section to this state is a factor of three smaller than the expected minimum one. The resonance position was varied so that the corresponding resonance energy $E_{\rm R}$ would take the values of 10, 100 and 500 keV, 
spanning the range of interest for BBN, for such a broad state. 
In these calculations, the Wigner limit value of 1.842 MeV was used for the reduced partial width in the entrance channel. 
This leads, at the three resonance energies mentioned above, to partial widths, $\Gamma_{\rm ^3He}$, of respectively 
5.70$\times$10$^{-43}$, 6.25$\times$10$^{-9}$ and 2.38 keV at a channel radius of 4.026 fm.  

Since  $\Gamma_{\rm ^3He}\ll\Gamma_{\rm T}$, the total width is dominated by the
exit channel partial widths, $\Gamma_{\rm T}$$\simeq$$\Gamma_{\rm p}$+$\Gamma_{\rm \alpha}$. 
To obtain the rates, we numerically integrated a Breit-Wigner formula assuming for the $^7$Be($^3$He,p)$^9$B 
and $^7$Be($^3$He,$\alpha$)$^6$Be reactions that $\Gamma_{\rm p}$=$\Gamma_{\rm T}$ and  $\Gamma_{\rm  \alpha}$=$\Gamma_{\rm T}$, respectively. 

The results are displayed in Fig.7 and correspond for each case to the  maximum possible rate. 
The calculated rates were included in a BBN nucleosynthesis calculation and were found to have no effect on the primordial $^7$Li/H abundance.
Thus, we can conclude that this channel is irrelevant for the $^7$Li problem. 

In summary, missing resonant states in the compound nuclei $^{10}$C and $^{11}$C nuclei were investigated 
via measurements at the Tandem-Alto facility of Orsay of the charge-exchange reactions $^{10}$B($^3$He,t)$^{10}$C and  
$^{11}$B($^3$He,t)$^{11}$C at the incident energy of 35 MeV. In the case of $^{10}$C, because of the presence of a large triton 
continuum background, the conclusion we could assert is that any state of $^{10}$C in the excitation energy region around 15 MeV requires 
a total width larger than 590 keV  to escape its detection in our experiment. With this experimental width limit and even with a three times lower one, 200 keV, the  $^7$Be+$^3$He calculated rates were found to be much too small to solve the lithium problem.  

Concerning the $^{11}$C case, the present data show that no new state is present at energies between 
7.79 and 7.9 MeV of $^{11}$C. 

\begin{figure}[h]
\begin{center}
\includegraphics[width=8.cm,height=7.6 cm]{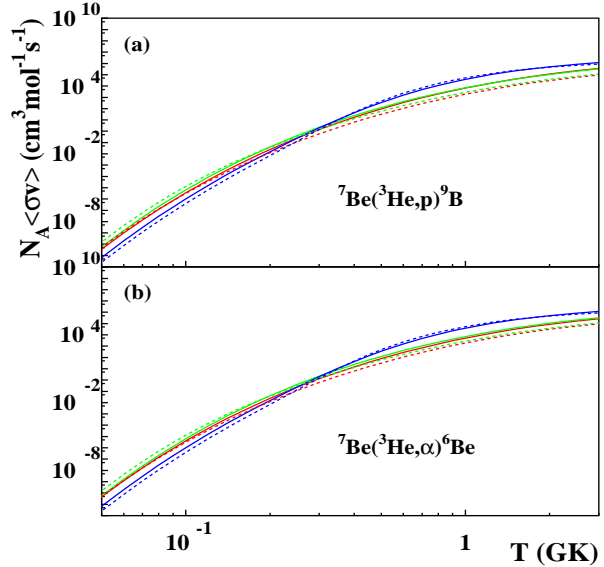}
\caption{(Color online) $^7$Be($^3$He,p)$^9$B and $^7$Be($^3$He,$\alpha$)$^6$Be reaction rate in (a) and (b) respectively. The red, green and blue solid lines correspond to calculations 
with $E_{R}$=10, 100 and 500 keV respectively and $\Gamma_{T}$=590 keV while the dotted lines correspond to calculations with $\Gamma_{T}$=200 keV. }
\end{center}
\end{figure}

Finally, our two results concerning $^{10}$C and $^{11}$C compound nuclei put an end to the various discussions concerning the missing resonant states in these nuclei  which were thought to partially or totally solve the $^7$Li problem \cite{CHA10,BRO12,CIV13} and exclude $^7$Be+$^3$He and $^7$Be+$^4$He  reaction channels as responsible for the observed $^7$Li deficit. With our conclusion and those of previous works concerning the other important reaction channels $^3$He+$^4$He and $^7$Be+d, the $^7$Li problem remains unsolved. 
The solution has very likely to be found outside of nuclear physics.

\acknowledgments

We thank the Tandem-Alto technical staff and the IPNO target laboratory staff
for their strong support during the experiment. 
This work has been supported by the European Community FP7 - Capacities-Integrated 
Infrastructure Initiative- contract ENSAR n$^\circ$ 262010,  the IN2P3-ASCR LEA NuAG 
and the French-Spanish AIC-D-2011-0820 project.


\begin{thebibliography}{99}

\bibitem{WMAP} E.~Komatsu et al.,  \zapjs, {\bf 192} 18 (2011). 

\bibitem{Planck} P. A. R. Ade et al. (The Planck Collaboration), arXiv:1303.5076, submitted to Astron. Astrophys. 

\bibitem{Cyb08} R.H.~Cyburt, B.O.~Fulop, and K.A.~Olive, J. Cosmology and Astroparticle Phys. 11, 12
(2008).

\bibitem{Coc10} A.~Coc and E.~Vangioni, Proc. {\it $4^{th}$ Int. Conf. on Nucl. Phys. in Astrophysics}, J. Phys. Conf. Ser. 202, 012001 (2010).

\bibitem{Coc13} A. Coc, J.-Ph. Uzan, E. Vangioni, arXiv:1307.6955 

\bibitem{Sbo10} L. Sbordone et al., Astron. Astrophys.,  522, 26 (2010)

\bibitem{Lithium2012a} M. Spite, F. Spite, and P. Bonifacio {\it Proc. "Lithium in the Cosmos" (Paris)}, {\it Mem. S.A.It.} {\bf 22} 9 (2012). 

\bibitem{Fie11} B. Fields, Annual Review of Nuclear and Particle Science, 61, 47 (2011)

\bibitem{Ric05} O.~Richard, G.~Michaud, \& J.~Richer, Astrophys. J. {\bf 619}, 538 (2005). 

\bibitem{Lithium2012b} F. Iocco, {\it Proc. "Lithium in the Cosmos" (Paris)}, {\it Mem. S.A.It.} {\bf 22} 19 (2012).

\bibitem{Coc04} A.~Coc, E.~Vangioni-Flam, P.~Descouvemont,
A.~Adahchour, and C.~Angulo, \zapj, {\bf 600} 544 (2004).

\bibitem{Ang05} C. Angulo  E.~Casarejos, M.~Couder et al., Astrophys. J.  {\bf 630}, L105 (2005). 

\bibitem{Cyb04} R. H. Cyburt, B. D. Fields and K. A. Olive, Phys. Rev. D 69, 123519 (2004)

\bibitem{DIL09} A. Di Leva et al. Phys. Rev. Lett 102, 232502 (2009) and references therein

\bibitem{Cyb12} R.H.~Cyburt \& M.~Pospelov,  {\it International Journal of Modern Physics} {\bf E21}, 50004 (2012), [arXiv:0906.4373].

\bibitem{OMa11} P.D.~O'Malley, D.W.~Bardayan, A.S.~Adekola et al.,  Phys. Rev.\ {\bf C84}, 042801 (2011).

\bibitem{Sch11} C.~Scholl, Y.~Fujita, T.~Adachi et al.,  Phys. Rev.\  {\bf C84}, 014308 (2011).

\bibitem{Kir11} O.S.~Kirsebom \& B.~Davids, Phys. Rev.\ {\bf C84}, 058801 (2011).

\bibitem{CHA10} N. Chakraborty, B. D. Fields and K. Olive Phys. Rev. D 83, 063006 (2011)

\bibitem{BRO12} C. Broggini, L. Canton, G. Fiorentini, F.L. Villante, JCAP 06, 030 (2012)

\bibitem{CIV13} O. Civitarese and M.E. Mosquera, Nucl. Phys. A 898, 1 (2013)

\bibitem{SPEN67} J. E. Spencer and H. A. Enge, Nucl. Instrum. Methods 49, 181 (1967)

\bibitem{SCHN75} M. J. Schneider et al. Phys. Rev. C 12, 335 (1975)

\bibitem{WAN93} L. Wang et al. Phys. Rev. C 47, 2123 (1993)

\bibitem{BEN67} W. Benenson, G. M. Crawley, J. D. Dreisbach and W. P. Johnson, Nucl. Phys. A 97, 510 (1967)  and references therein

\bibitem{NNDC} http://www.nndc.bnl.gov/chart/ and references therein

\bibitem{Ball67} G. C. Ball et al. Phys. Rev. 155, 1170 (1967)

\bibitem{Ball69} G. C. Ball and J. Cerny, Phys. Rev. 177, 1466 (1969)

\bibitem{STE83} W. A. Sterrenburg, M. N. Harakeh, S. Y. Van Der Werf and Van Der Woude, Nucl. Phys. A 405, 109 (1983).



\end{thebibliography}
\end{document}